\documentclass[showpacs, pra,onecolumn,preprintnumbers ,amsmath, amssymb, superscriptaddress, aps]{revtex4-2}
\usepackage{amsfonts}
\usepackage{color}
\usepackage{amsmath,amssymb}
\usepackage{pifont}
\usepackage{amssymb}  
\usepackage{bbold}
\usepackage{float}
\usepackage{subfloat}

\usepackage[caption=false]{subfig}
\usepackage{tikz}
\usepackage{makecell}
\usepackage{subfig}
\usepackage{pifont}   
\usepackage{graphicx} 
\graphicspath{{Figures/}}
\usepackage{dcolumn}  
\usepackage{bm}       
\usepackage{multirow} 
\usepackage{placeins}
\usepackage[colorlinks]{hyperref}
\usepackage{mathtools}
\usepackage{appendix}

\def \be{\begin{align}}
	\def \ee{\end{align}}
\def \bea{\begin{eqnarray}}
	\def \eea{\end{eqnarray}}

\begin{document}
	
	\title{Effects of AB-flux  and gap on magnetic graphene  quantum dots}
	
	\author{Mohammed El Azar}
		\email{elazar.m@ucd.ac.ma}
	\affiliation{ Laboratory of Theoretical Physics, Faculty of Sciences, Choua\"ib Doukkali University, PO Box 20, 24000 El Jadida, Morocco}
	\author{Ahmed Bouhlal}
	\affiliation{ Laboratory of Theoretical Physics, Faculty of Sciences, Choua\"ib Doukkali University, PO Box 20, 24000 El Jadida, Morocco}
	\author{Ahmed Jellal}
	\email{a.jellal@ucd.ac.ma}
	\affiliation{ Laboratory of Theoretical Physics, Faculty of Sciences, Choua\"ib Doukkali University, PO Box 20, 24000 El Jadida, Morocco}
	\affiliation{
		Canadian Quantum  Research Center,
		204-3002 32 Ave Vernon, BC V1T 2L7,  Canada}

	\begin{abstract}
		We consider magnetic graphene quantum dots (MGQDs) and study the impact of the Aharonov-Bohm (AB) flux and gap on the scattering process of electrons. Our emphasis is on the finite lifetimes of quasi-bound states arising from the interaction between electrons and the magnetic field within the dot. Initially, we calculate the scattering coefficients, scattering efficiency, and probability density by ensuring the continuity of eigenspinors at the boundary of MGQD. The results indicate that as the  gap increases, the quasi-bound states reach higher maxima. We show that an increase in AB-flux leads to a generation of quasi-bound states requiring less magnetic field, and the scattering efficiency starts to take non-zero values at smaller MGQD sizes. The analysis of probability density shows that the quasi-bound states, corresponding to non-resonantly excited scattering modes, exhibit a significant improvement in the concentrated density at MGQD. The improvement is a result of reducing the diffraction phenomenon and suppressing the Klein effect through an increase in AB-flux and  gap. This increases the probability of retaining the electron for a longer period of time.
	
	\vspace{0.25cm}
	
\end{abstract}
\maketitle

\section{Introduction}
Graphene quantum dots (GQDs) are indeed nanoparticles, typically smaller than 100 nanometers in size \cite{Than2022,Cui2024}. GQDs are small pieces of graphene with unique electronic and optical properties due to their quantum confinement effect. They have gained significant attention in various fields, including electronics, optoelectronics, energy storage, and biomedicine \cite{Haque2018,  Kalluri2018, Chung2021, Balak2023}.
Furthermore, GQDs have been the subject of intense research in recent years due to their unique electronic and optical properties \cite{Zarenia11, Bischoff15, Krivenkov17, Choi17, Sadrara19, Bouhlal23}. Apart from their size and properties, the GQDs behave like a potential well, confining electrons (holes) in space. This confinement occurs within a region of the electron's wavelength. GQDs have properties similar to those of atoms, sometimes called artificial atoms \cite{Kastner93}, and can be considered very small pieces of graphene \cite{Zhu15}. Other theoretical aspects show that GQDs are promising candidates for future quantum information techniques. Several studies have explored the possibility of creating GQDs to trap electrons within them \cite{Lee16, Gutierrez16}. On the other hand, it is experimentally observed that the manifestation Klein tunneling in graphene \cite{Katsnelson06}. Initially, this tunneling was demonstrated by Klein in 1929 \cite{Klein29}, which is a relativistic phenomenon where electrons can tunnel through a potential barrier with 100 $\%$ efficiency, regardless of the barrier height. 
It is not possible to permanently localize an electron within a graphene system due to Klein tunneling. However, some publications suggest that it is possible to trap or confine electrons in graphene space, such as quantum dots, despite the Klein tunneling phenomenon \cite{Beenakker08, Allain11, Bouhlal21}.


Short-term electron trapping in GQDs under external electrostatic potentials was investigated \cite{Fehske15, Lee16}. Electronic states, known as quasi-bound states, differ from truly bound states, such as those in an atom, and are typically characterized by a finite lifetime, commonly referred to as trapping time.
By analogy, the temporary presence of an electron inside the GQD is referred to as a quasi-localization, which disappears due to the Klein tunneling effect, which is responsible for escaping electrons from the GQD.
Trapping electrons in materials for a period of time is necessary to control their mobility. However, this is challenging in graphene due to its zero energy gap and Klein tunneling. Despite this, some studies have suggested the possibility of temporary electron trapping. 
Moreover, twisted light \cite{Pena22}, magnetic fields \cite{Martino07, Wang09, Pan20}, and polarized light \cite{Penaa22} can be employed to generate quasi-bound states in GQDs, extending the trapping time of electrons.




Recently, we conducted two different studies on how external sources affect electron trapping in GQDs. The first study explored the effect of the mass term \cite{Azar23}, and the second focused on the impact of the Aharnov-Bom (AB) flux \cite{Azar24} on GQDs. Here, we combine both works to treat the influence of AB-flux and gap on the scattering process in magnetic graphene quantum dots (MGQDs). 
 We initiate the process by analytically solving the Dirac equation to determine the corresponding eigenspinors. Subsequently, we apply the boundary conditions to obtain the basic formulas for the primary physical variables needed to analyze scattering phenomena, including the scattering coefficient, scattering efficiency, and probability density. 
 To investigate the lifetime of quasi-bound states, we set up an equation  in the complex space of incident energies. The results indicate that as the energy gap widens, the quasi-bound states reach higher peaks.
 We note that the scattering efficiency starts to exhibit non-zero values at smaller MGQD sizes. Additionally, we demonstrate that a higher AB-flux creates quasi-bound states that demand less magnetic field.
We demonstrate that the probability density reveals suppression of diffraction and Klein tunneling for excited quasi-bound states as AB-flux and gap increase.
This accounts for the notable enhancement in the probability density concentrated within the MGQD, leading to an increased probability of retaining the electron for an extended period.


The paper is organized as follows:  Sec. \ref{theory} introduces a theoretical model that accurately describes the scattering process. We solve the Dirac equation to identify the eigenspinors in different regions of the system. In Sec. \ref{Scettering}, we apply the boundary condition to determine the analytical expressions of various physical quantities associated with the diffusion phenomenon. These include the scattering coefficients, the scattering efficiency, and the probability density. Sec. \ref{res} presents a numerical analysis of the theoretical results, considering various values of the physical quantities that characterize the system. These include incident energy, AB-flux, quantum dot radius, gap, and magnetic field intensity. Finally, in Sec. \ref{cc}, we provide a brief conclusion.

\section{Theoretical model}\label{theory}

\begin{figure}[h]
	\centering 
	\includegraphics[scale=0.3]{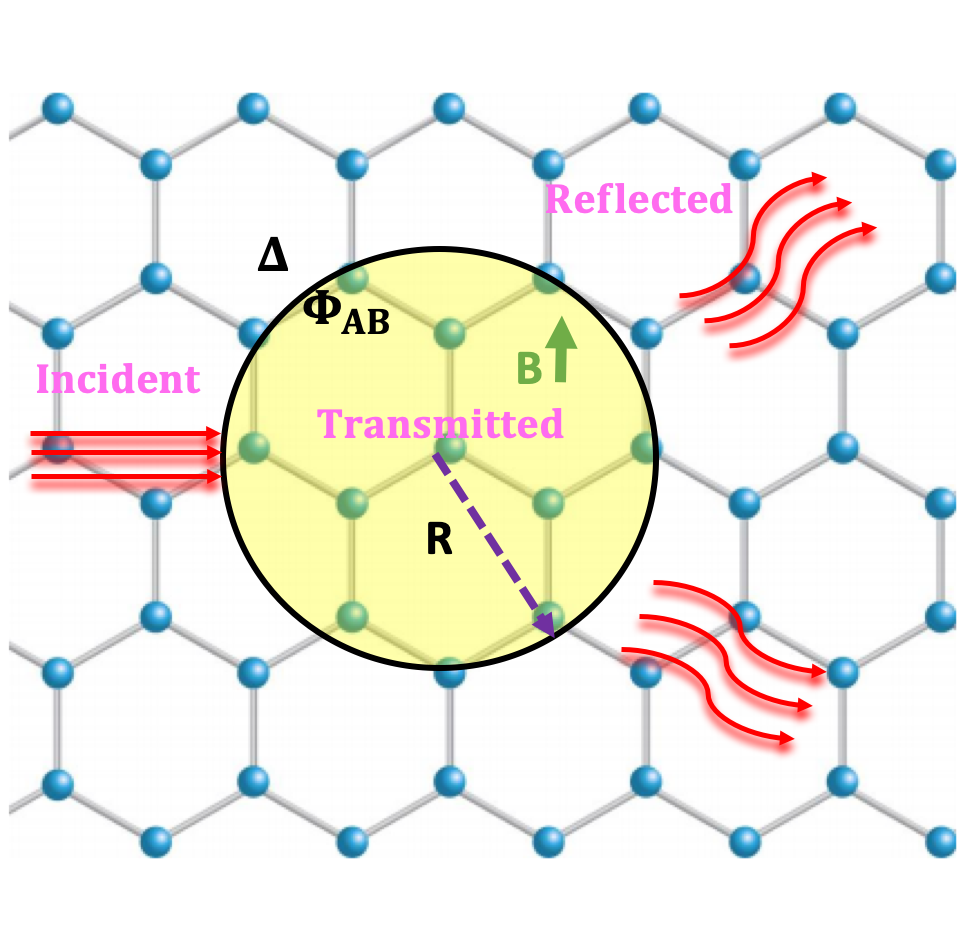}	
	\caption{(color online) MGQD of radius $R$ is confined by a constant magnetic field $B$ and exposed to an AB-flux $\Phi_{AB}$  inside and a gap $\Delta$ outside.}\label{system}
\end{figure}

Figure \ref{system} shows the fundamental characteristics of the system under study. This system can be simulated by a MGQD of radius $R$ where the confinement is ensured by a magnetic field $B$, an AB-flux $\Phi_{AB}$ is applied inside, and a gap $\Delta$ is outside. An electron of energy $E$ at normal incidence is dispersed by the plane wave $\psi_i^k$. The wave function $\psi_r^k$ represents the reflected component, and $\psi_t^q$ represents the transmitted one. 
We choose $\Delta$ and $\vec{B}$ as 
\begin{equation}
\Delta(r)= \begin{cases} \Delta , & r>R \\ 0, & \text { otherwise }\end{cases}, \quad \vec{B}= \begin{cases} B \vec{e}_z, & r<R \\ 0, & \text { otherwise }\end{cases}
\end{equation} 
where $\vec{e}_z$ is the unit vector along the $z$-direction. The Hamiltonian governing our system can be expressed as 
 \begin{equation}\label{Hamilt}
 H =v_F  (\vec p +e \vec A)\cdot \vec\sigma+ \Delta \sigma_z.
 \end{equation}
where $v_F = 10^6$ ms$^{-1}$ is the Fermi velocity, $\vec p$ is the momentum operator, $\sigma=(\sigma_x,\sigma_y,\sigma_z)$ are  Pauli matrices, and $(-e)$
is  the elementary charge. The vector potential $\vec{A}$ is expressed as the sum of two components: the first $\vec{A_1}=\frac{Br}{2} \vec{e}_\varphi$ is related to the applied magnetic field, and the second $\vec{A_2}=\frac{\Phi_{AB}}{2\pi r}\vec{e}_\varphi$ is related to the applied extra AB-flux field \cite{Ikhdair15}. Therefore, in gauge symmetry, the vector potential can be defined as 
\begin{equation}
\vec{A}= \begin{cases} \left( \frac{Br}{2}+\frac{\Phi_{AB}}{2\pi r}\right)\vec{e}_\varphi , & r<R \\ 0, & \text { otherwise }.\end{cases}
\end{equation} 
Because of the spherical symmetry, we can work within the polar representation $(r, \varphi)$. Then, we express the Hamiltonian \eqref{Hamilt} as
\begin{equation}\label{ham3}
H =\begin{pmatrix} \Delta & -i \hbar v_F  e^{-i\varphi}\left[\frac{d}{dr} - \frac{i}{r}\frac{d}{d\varphi} -\frac{er}{2\hbar}\left(B+\frac{\Phi_{AB}}{\pi r^2}\right) \right]\\\\
-i \hbar v_F e^{i\varphi}\left[\frac{d}{dr} + \frac{i}{r}\frac{d}{d\varphi}+ \frac{er}{2\hbar}\left(B+\frac{\Phi_{AB}}{\pi r^2}\right) \right]	&-\Delta \\
\end{pmatrix}	.
\end{equation}		
Since the Hamiltonian \eqref{ham3} commutes with the total angular momentum $J _z= -i\hbar \partial_ \varphi +\frac{\hbar}{2}\sigma_z$, then we look for the eigenspinors that form a common basis for $H$ and $J_z$. They are
\begin{equation}\label{ansatz}
\psi(r,\varphi)= e^{im\varphi}\dbinom{ \psi_A(r)}{ie^{i\varphi}\psi_B(r)}
\end{equation}
where the quantum numbers $m \in \mathbb{Z}$ denote the eigenvalues of $J_z$. 

In the forthcoming analysis, we concentrate on the elastic scattering process of electrons in the MGQD of radius $R$ under consideration. It is worth noting that an electron moving freely in the $x$-direction in the absence of a magnetic field will fall onto the MGQD with an energy $E$, where $k$ is a modulo of the wave vector $\vec{k}$. This electron is actually described by the incident spinor
\begin{align}
\psi_i^k(r, \varphi) =\frac{1}{\sqrt{2}} e^{ikr\cos\varphi}\dbinom{1}{1}=\frac{1}{\sqrt{2}} \sum_{m=-\infty}^{\infty} i^m \dbinom
{e^{i m\varphi} J_m(k r)} 
{i e^{i(m+1) \varphi} J_{m+1}(k r)}.
\end{align}
and we have 
$k=\frac{\sqrt{E^2-\Delta^2}}{\hbar v_F}$
 with  $J_m(x)$ are Bessel functions of the first kind. The incident wave, when it is striking the MGQD, is therefore either transmitted or reflected, and each of these latter cases is characterized by a specific probability. The reflected wave is assumed to be a decomposition into partial waves \cite{Heinisch13, Schulz15}
\begin{align}
\psi_r^k(r, \varphi) =\frac{1}{\sqrt{2}} \sum_{m=-\infty}^{\infty} i^m \Gamma_r^m\dbinom
{e^{i m\varphi} H_m(k r)} 
{i e^{i(m+1) \varphi} H_{m+1}(k r)}.
\end{align}
where $H_m (x)$ are Hankel functions of the first kind, and $\Gamma_r^m$ are reflection coefficient, which can be determined by considering   the asymptotic behavior 
\begin{align}
	H_m(x) \underset{x\gg 1}{\sim} \sqrt{\frac{2}{\pi x}} e^{i(x-\frac{m\pi}{2}-\frac{\pi}{4})}.
\end{align}

The solution inside ($r<R$) the MGQD with the presence of  magnetic field $B$ and  AB-flux  $\Phi_{AB}$ can be obtained by solving the Dirac equation $H\psi^q(r,\varphi) = E\psi^q(r, \varphi)$ where $ E =\pm \hbar v_F q$ and the signs $ \pm$ differentiating between the valence and conduction bands. As a result, we arrive at the following two coupled equations
\begin{subequations}\label{8}
	\begin{align}
	&		\frac{d}{dr}\psi_A^q(r)	+\left( \frac{r}{2 l_B^2}-\frac{\beta}{r} \right) \psi_A^q(r) =-q \psi_B^q(r) \label{8a}\\
	&  \frac{d}{dr}\psi_B^q(r)	-\left( \frac{r}{2l_B^2}-\frac{\beta+1}{r} \right) \psi_B^q(r) = q \psi_A^q(r) \label{8b}
	\end{align}
\end{subequations}	
where $q$ is the wave number associated with the electron inside the MGQD, $l_B=\sqrt{\frac{\hbar}{e B}}$ is the magnetic length, we have set the parameter 
$\beta= m-\phi$, with the rescaled AB-flux $\phi=\frac{\Phi_{AB}}{\phi_0}$ and the quantum unit $\phi_0=\frac{h}{e}$. By injecting \eqref{8a} into \eqref {8b}, we get a second-order differential equation for $ \psi_ A^ q(r) $
\begin{equation}\label{e:9}
\frac{d^2}{dr^2}\psi_A^q(r)+ \frac{1}{r}\frac{d}{dr}\psi_A^q(r) + \left( \frac{\beta+1}{l_B^2}-\frac{r^2}{4 l_B^4}-\frac{\beta^2}{r^2}+q^2\right)\psi_A^q(r) =0.
\end{equation}
In order to find an approximate solution, we exploit the asymptotic limits that resemble the physical behavior sufficient to describe our system. Then, in limits $r\rightarrow 0$ and $r\rightarrow \infty$, \eqref{e:9} reduces to two forms 
\begin{subequations}\label{e:10}
	\begin{align}
	&		\frac{d^2}{dr^2}\psi_A^q(r)+ \frac{1}{r}\frac{d}{dr}\psi_A^q(r) -\frac{\beta^2}{r^2}\psi_A^q(r) =0 \label{10a}\\
	& \frac{d^2}{dr^2}\psi_A^q(r)+ \frac{1}{r}\frac{d}{dr}\psi_A^q(r) -\frac{r^2}{4 l_B^4}\psi_A^q(r) =0 \label{10b}.
	\end{align}
\end{subequations}	
having the following solutions 
\begin{subequations}\label{e:11}
	\begin{align}
	&		\psi_A^q(r) =\frac{c_1}{2}(r^\beta +r^{-\beta})+\frac{i c_2}{2}(r^\beta -r^{-\beta}) \label{11a}\\
	& 	\psi_A^q (r) =c_3 I_0\left( \frac{r^2}{4l_B^2}\right) +c_4 K_0\left( \frac{r^2}{4l_B^2}\right)  \label{11b}
	\end{align}
\end{subequations}	
where $I_0(x)$ and $K_0(x)$ denote the zero-order modified Bessel functions of the first and second kinds, respectively. By considering the asymptotic behavior of  $K_0(x) \underset{x\gg 1}{\sim} \frac{e^{-x}}{\sqrt{x}}$ and   choosing the constants ($c_1, c_2, c_3, c_4$) \cite{Pena222}, we show that  \eqref{e:11} gives 
	\begin{subequations}\label{e:12}
	\begin{align}
	&\psi_A^q(r) \sim\begin{cases} r^\beta ,& \beta\ge 0 \\
	r^{-\beta},& \beta<0 \label{12b} 
	\end{cases}\\
	&\psi_A^q(r) \sim 2 l_B \frac{e^{-r^2/4 l_B^2}}{r}.\label{12c}
	\end{align}
\end{subequations}
Consequently, a general solution of   \eqref{e:9} takes the form 
	\begin{equation}\label{e:13}
\psi_{A\pm}^q (r) =2 l_B r^{\pm \beta }  \frac{e^{-r^2/4 l_B^2}}{r} \Omega_{A\pm}^q(r).
\end{equation}
Now using the variable change  $s=\frac{r^2}{2 l_B^2} $ and the transformation $ \Omega_{A\pm}^q(s)=\sqrt{s} \Sigma_{A\pm}^q(s)$ to transform \eqref{e:9} into  the two Kummer-type differential equations for $\Sigma_{A+}^q(s)$ and $\Sigma_{A-}^q(s)$
\begin{subequations}\label{e:14}
	\begin{align}
	&s \frac{d}{ds^2} \Sigma_{A+}^q(s) +\left( \beta+1-s\right)  \frac{d}{ds} \Sigma_{A+}^q(s) +\frac{l_B^2 q^2}{2} \Sigma_{A+}^q(s)=0;  \label{14a}\\
	&s \frac{d}{ds^2} \Sigma_{A-}^q(s) +\left( 1-\beta-s\right)  \frac{d}{ds} \Sigma_{A-}^q(s)
	+\left( \beta+\frac{l_B^2 q^2}{2}\right) \Sigma_{A-}^q(s)=0 \label{14b} 
	\end{align}
\end{subequations}
and the corresponding solutions are confluent hypergeometric functions as listed below
\begin{subequations}\label{15}
	\begin{align}
	&\Sigma_{A+}^q(s)=\prescript{}{1}{F}_1^{}\left(-\frac{l_B^2 q^2}{2},\beta+1,s\right) \label{15a}\\
	&\Sigma_{A-}^q(s)=\prescript{}{1}{F}_1^{}\left(-\beta-\frac{l_B^2 q^2}{2},1-\beta,s\right). \label{15b} 
	\end{align}
\end{subequations}
Therefore, we end up with 
\begin{subequations}\label{16}
	\begin{align}
	& \psi_{A+}^q (s)= s^{\vert \beta \vert / 2} e^{-s / 2}\prescript{}{1} F_1\left(-\frac{l_B^2 q^2}{2}, \beta+1, s\right); \label{16a}\\
	& \psi_{A-}^q (s)= s^{\vert \beta \vert / 2} e^{-s / 2} \prescript{}{1}F_1 \left(-\beta-\frac{l_B^2 q^2}{2},1-\beta, s\right). \label{16b} 
	\end{align}	
\end{subequations}	
By injecting  \eqref{16a} and \eqref{16b}, respectively, into \eqref{8a}, we establish  the remaining components  $\psi_{B\pm}^q (s)$
\begin{subequations}
	\begin{align}
	& \psi_{B+}^q (s)= \frac{q l_B / \sqrt{2}}{\vert \beta \vert+1} s^{(\vert \beta \vert+1) / 2} e^{-s / 2}\prescript{}{1} F_1\left(1-\frac{l_B^2 q^2}{2}, \beta+2, s\right)	\\
	& \psi_{B-}^q (s)=-\frac{\vert \beta \vert}{q l_B / \sqrt{2}} s^{(\vert \beta \vert -1)/ 2} e^{-s / 2} \prescript{}{1}F_1 \left(-\beta-\frac{l_B^2 q^2}{2},-\beta, s\right).
	\end{align}	
\end{subequations}	
Finally, we combine all the obtained results to write the eigenspinors inside the MGQD in the following manner
\begin{equation}\label{17}
\psi_t^q(r,\varphi) =\sum_{m=-\infty}^{\infty} \Gamma_t^m  \dbinom{\psi_{A\pm}^q (r) e^{im\varphi}}{i \psi_{B\pm}^q(r)e^{i(m+1)\varphi}}
\end{equation}
where $\Gamma_t^m $ are transmission coefficients. The above solutions will be employed to discuss the scattering process for the present system.

\section{Scattering quantities}\label{Scettering}

To characterize the scattering phenomena of our system, we need to determine the physical characteristics, including the scattering efficiency $Q$, the scattering coefficients $\Gamma_r^m$ and $\Gamma_t^m$, and the probability density $\rho$. Indeed, we use the continuity of the eigenspinors at the boundary of the MGQD
\begin{align}	
\psi_i^k(R,\varphi) +\psi_r^k(R,\varphi) =\psi_t^q(R,\varphi)
\end{align}
 to establish  two coupled equations for $\Gamma_r^m$ and $\Gamma_t^m$
\begin{subequations}\label{18}
	\begin{align}
	& i^m J_m(kR)+ i^m \Gamma_r^m H_m(kR) = \sqrt{2} \Gamma_t^m \psi_{A\pm}^ q(R) \label{18a}\\
	&	 i^m J_{m+1}(kR)+ i^m \Gamma_r^m H_{m+1}(kR) = \sqrt{2} \Gamma_t^m \psi_{B\pm}^q (R).  \label{18b} 
	\end{align}
\end{subequations}
By involving the relation between the first kind Bessel and Hankel functions
\begin{align}
	 H_m(x) J_{m+1}(x) - H_{m+1}(x) J_m(x) =\frac{2i}{\pi x}
\end{align}
  and after some algebra, we get 
\begin{subequations}\label{19}
	\begin{align}
	\Gamma_{t\pm}^m&=\frac{i^m} {\sqrt{2}}\  \frac{J_m(kR) H_{m+1}(kR)- J_{m+1}(kR) H_{m}(kR)}{
		H_{m+1}(kR)\psi_{A\pm}^ q(R)-H_{m}(kR)
		\psi_{B\pm}^q (R)} \label{19a}\\
	\Gamma_{r\pm}^m&=\frac{-J_m(kR)\psi_{B\pm}^q (R)+J_{m
			+1}(kR)\psi_{A\pm}^q (R)}{H_m(kR)\psi_{B\pm}^q (R)-H_{m
			+1}(kR)\psi_{A\pm}^q(R) }.  \label{19b} 
	\end{align}
\end{subequations}
As for the probability density $\rho$, we consider the definition
\begin{equation}\label{20}
 \rho= \psi^\dag \psi.
\end{equation}
where $\psi$ can be substituted  by $\psi_i^k+\psi_r^k$ outside,  and $\psi_t^q$ inside the quantum dot. 
In order to examine how the size of the quantum dot affects the scattering phenomenon, we introduce the scattering efficiency. It is defined as the ratio of the scattering cross-section to the geometrical cross-section \cite{Schulz15, Belokda22}
\begin{equation}\label{21}
Q=\frac{\sigma}{2 R}=\frac{4}{k R}\sum_{m=-\infty}^{+\infty} \vert \Gamma_{r\pm}^m(\phi)\vert^2.
\end{equation}

It is well known that the quasi-bound states have  positive energy on the continuum. Now if we want to study the scattering phenomenon in a precise way, taking into account the lifetime of the quasi-bound states, we have to go into the complex space of energies of the incident electron \cite{Narimanov99}
\begin{align}
	E=E_r-iE_i
\end{align}
where the real component $E_r$ of the complex energy signifies the resonance energy, and $E_i$ is a real quantity linked to the lifetime (trapping time)   by $\tau=\frac{\hbar}{E_i}$. We can also use the $k$-representation, knowing that $E=\pm \hbar v_F k$, with $k = k_r - ik_i$, to express $\tau$ as
\begin{equation}\label{22}
\tau=\frac{1}{v_F k_i}.
\end{equation}
We determine the complex energy of the incident electron by equating the transmitted wave function to the reflected wave function at the boundary $r = R$  \cite{Hewageegana08}. As the incident energy of the incoming electron remains unaffected by the magnetic field, we address the subsequent transcendental equation for $q$ and $k$
%
\begin{equation}\label{e:34}
\frac{\psi_{A\pm}^q(R)}{\psi_{B\pm}^ q(R)}=\frac{H_m(kR)}{H_{m+1}(kR)}.
\end{equation} 
The numerical resolution of this equation allows us to establish resonances, consequently determining a set of incident energies $(E_r, E_i)$ associated with the magnetic field $B$. Consequently, we can identify the lifetime $\tau$ of each quasi-bound state.

\section{RESULTS AND DISCUSSIONS}\label{res}
\begin{figure}[ht]
	\centering
\includegraphics[scale=0.46]{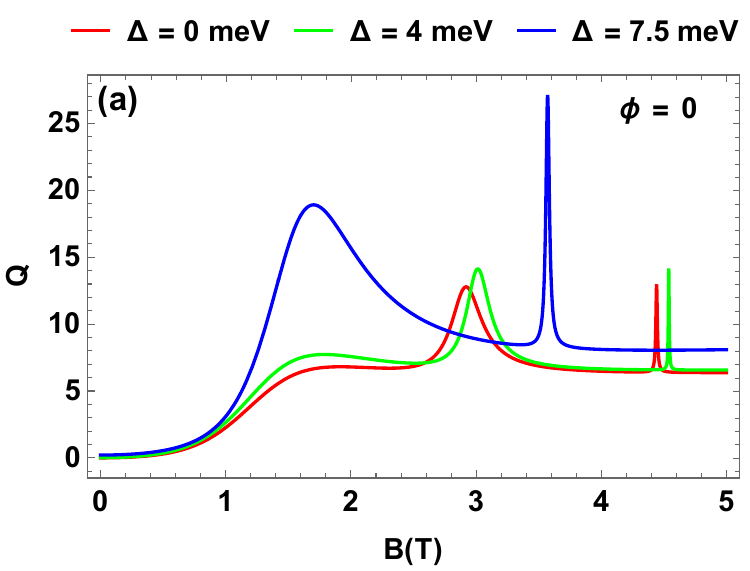}  \includegraphics[scale=0.46]{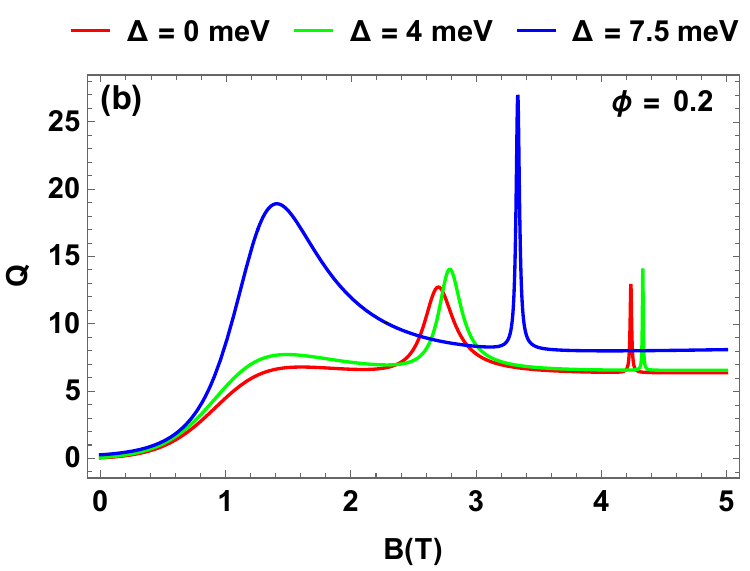}  \includegraphics[scale=0.46]{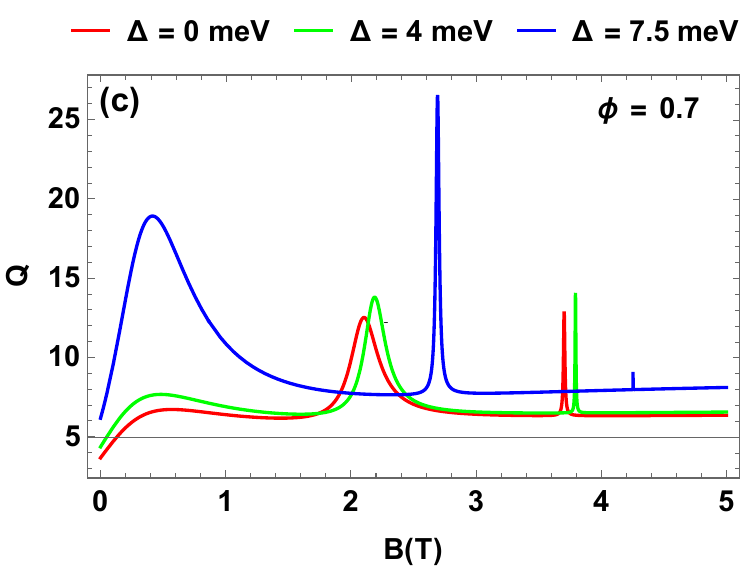}
\includegraphics[scale=0.46]{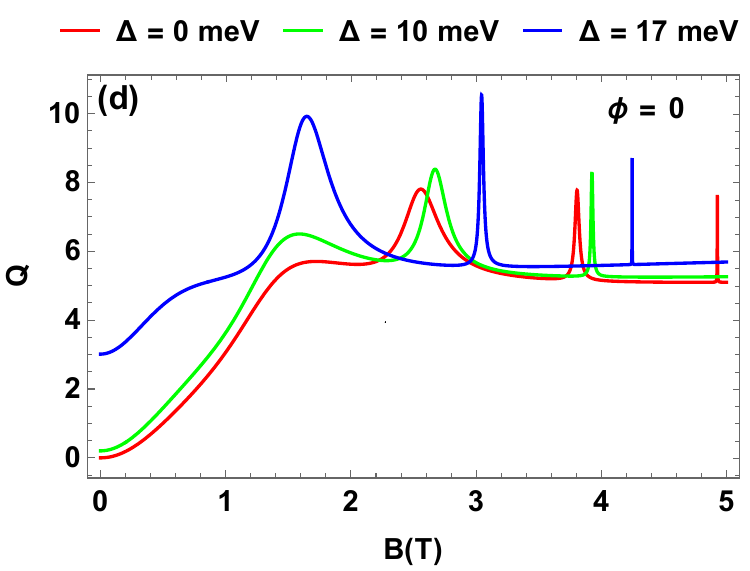}\includegraphics[scale=0.46]{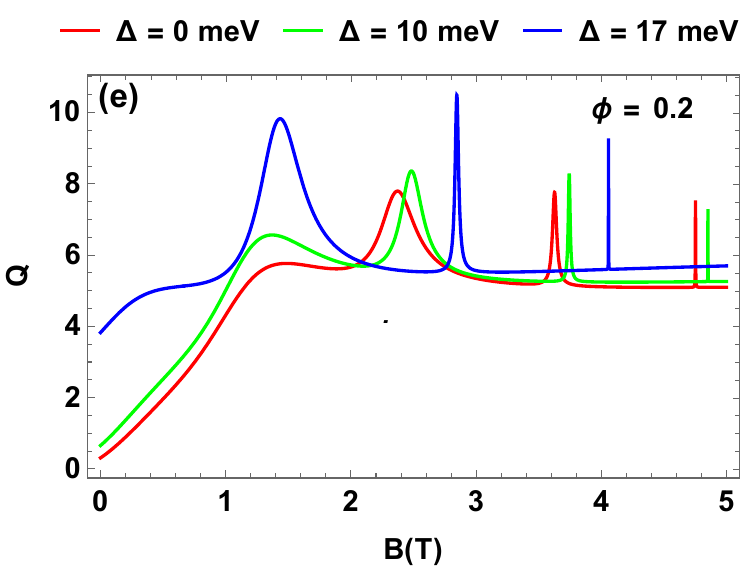}\includegraphics[scale=0.46]{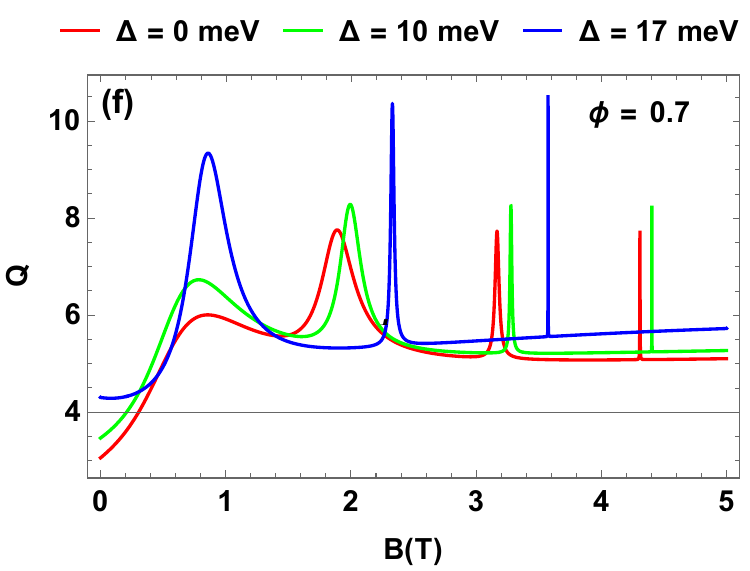}
	\caption{(color online) Scattering efficiency $Q$ as a function of magnetic field  $B$ for radius $R=50$ nm, three rescaled AB-flux values $\phi = 0, 0.2, 0.7$, two incident energy cases (a,b,c): $E=8$ meV,  (d,e,f): $E=20$ meV, along with different values of  gap $\Delta$.} 
	\label{fig2art}
\end{figure}

To delve more profoundly into the properties of the present system and glean further insights from our theoretical findings, we suggest undertaking a numerical analysis. By employing numerical methods under different configurations of the physical parameters, we can explore the behavior of the system in finer detail and shed light on its fundamental characteristics. This approach enables us to extract valuable information and enhance our understanding the dynamics and behavior of the system.
In Fig. \ref{fig2art}, we plot the scattering efficiency $Q$  as a function of the magnetic field  $B$ for a MGQD of radius $R = 50$ nm with different values of incident energy $E$, rescaled AB-flux $\phi$, and gap $\Delta$.  In the absence of flux ($\phi=0$), Fig. \ref{fig2art}a illustrates that $Q$ only starts to assume non-null values when $B$ reaches a value close to 1 T.  
Conversely, we observe that as $\Delta$ increases, the resonance peaks shift slightly towards larger values of $B$, and $Q$ attains higher values. In Figs. \ref{fig2art}(b,c), it is evident that even in the absence of $B$, $Q$ takes non-null values as long as the flux is present inside the MGQD. We observe that the resonance peaks corresponding to the excited scattering modes shift significantly to the left (with smaller values of $B$) as $\phi$ increases.
The findings presented in \cite{Azar24} provide further support for the conclusions we have discussed. In Figs. \ref{fig2art}(d,e,f) for the energy $E=20$ meV, it is clear that the amplitude of the resonance peaks consistently increases with the rise of $\Delta$. Additionally, we note that the greater the increase in $\phi$, the smaller the values of $B$ required for the generation of quasi-bound states. It is crucial to emphasize that for an incident electron energy of $E=20$ meV, $Q$ exhibits a noticeable enhancement. This improvement can be attributed to the apparent enhancement in the minimum value of $Q$ at $B=0$ and the increased number of excited scattering modes compared to the case of $E=8$ meV. Therefore, our analysis below will focus on the incident energy $E=20$ meV.



\begin{figure}[ht]
	\centering
	\includegraphics[scale=0.42]{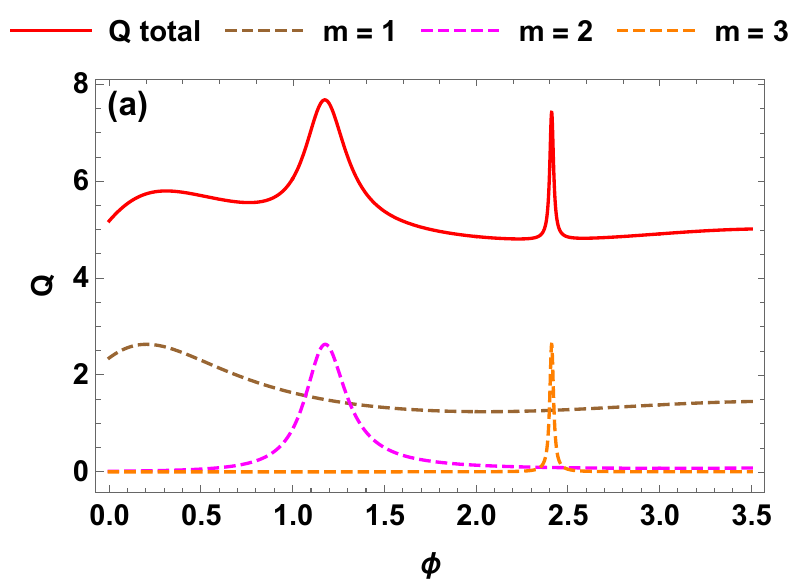}  \includegraphics[scale=0.42]{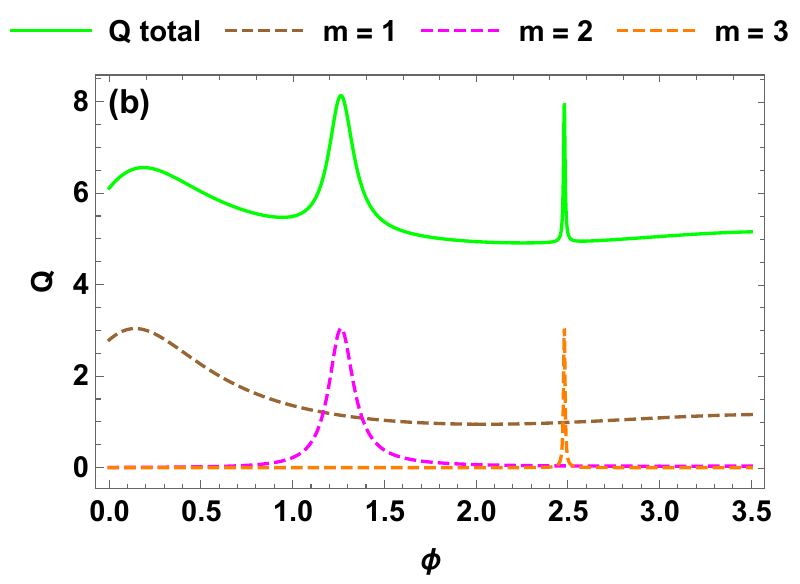}  \includegraphics[scale=0.42]{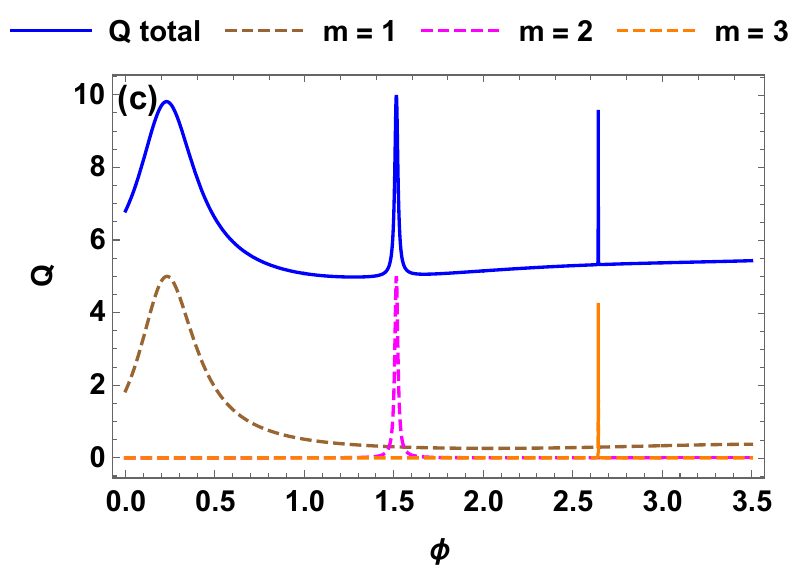}
	\caption{(color online) Scattering efficiency $Q$ as a function of the rescaled AB-flux $\phi$  for radius $R=50$ nm,  incident energy $E=20$ meV, magnetic field  $B= 1.4$ T, modes $m=1,2,3$, along with (a): $\Delta=0$, (b): $\Delta=10$ meV, (b): $\Delta=17$ meV. }
	\label{fig3art}
\end{figure}

Fig. \ref{fig3art} presents the scattering efficiency $Q$ as a function of the rescaled AB-flux  $\phi$  for  $E=20$ meV, $B=1.4$ T, $R=50$ nm, three scattering modes $m=1,2,3$, and three gap values $\Delta=0,10,17$ meV. In Fig. \ref{fig3art}(a) for $\Delta=0$, we provide two observations. Firstly, $Q$ has a notable minimum value in the absence of AB-flux ($\phi=0$). Secondly, $Q$ exhibits an oscillatory behavior with three peaks corresponding to the three scattering modes $m=1,2,3$. To elaborate on our statement, we will discuss each mode individually. For $m=1$, $Q$ is excited at $\phi \approx 0$ in a non-resonant manner, and therefore, we do not anticipate any significant trapping effect. For $m=2$, $Q$ is excited at $\phi=1.17$, but its peak does not exhibit complete resonance, leading to the emergence of a trapping effect. Since the $m=3$ mode is generated resonantly at $\phi=2.4$, we expect an observable trapping effect. As $\Delta$ increases, the peaks corresponding to the three scattering modes become narrower and shift towards higher values of the AB-flux. This has an adverse impact on the trapping effect, especially for the mode $m=3$, as depicted in Figs. \ref{fig3art}(b,c).

%



	\begin{figure}[H]
	\centering
	\includegraphics[scale=0.36]{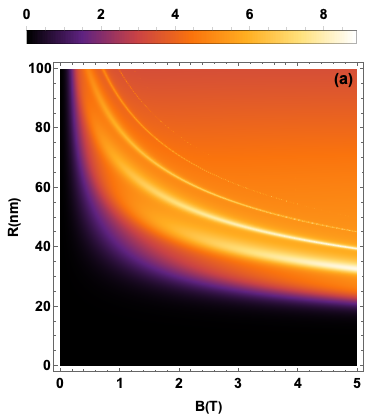}  \includegraphics[scale=0.36]{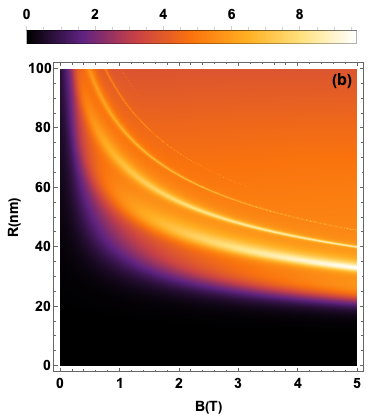}  \includegraphics[scale=0.36]{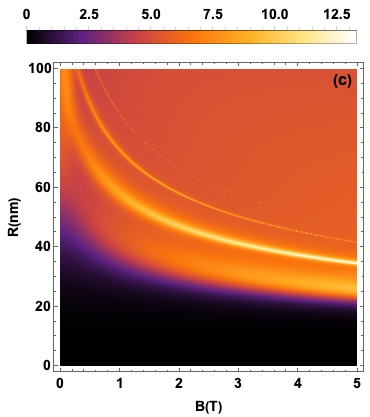}
	 \includegraphics[scale=0.36]{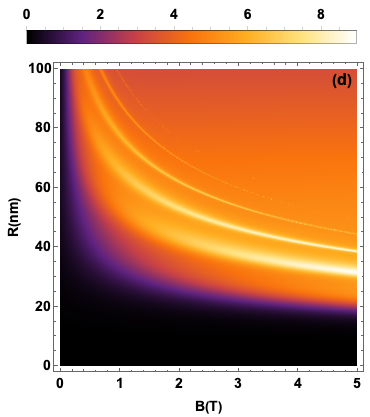} \includegraphics[scale=0.36]{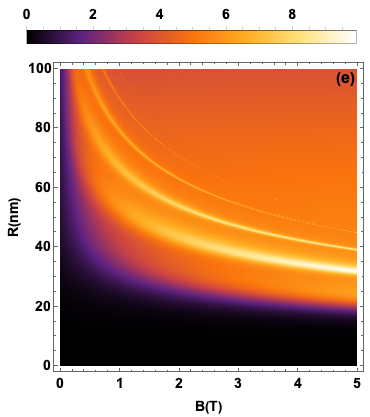}  \includegraphics[scale=0.36]{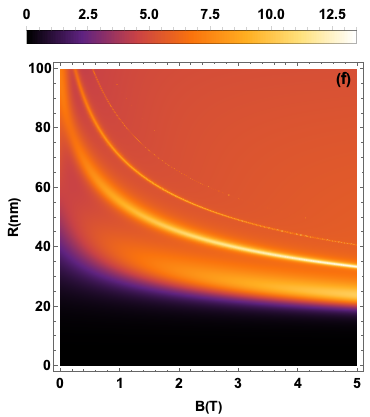}  
	 \includegraphics[scale=0.36]{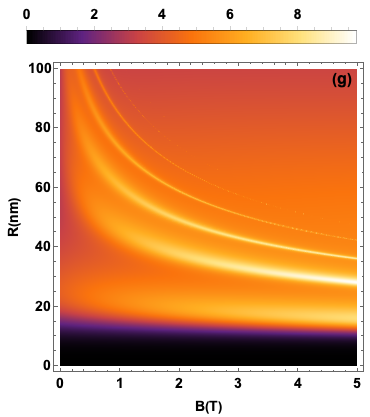} \includegraphics[scale=0.36]{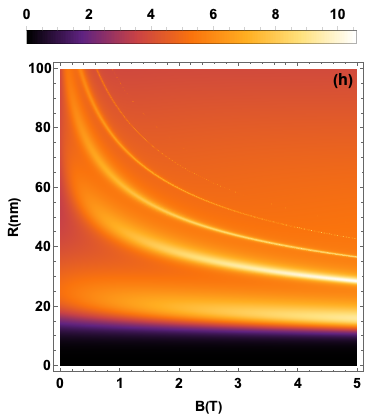} \includegraphics[scale=0.36]{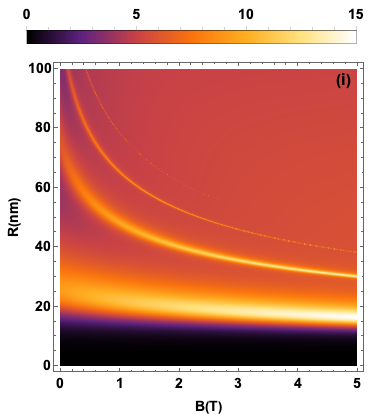} 
	\caption{(color online) Scattering efficiency $Q$ as a function of the radius $R$ and magnetic field $B$ for  $E=20$ meV, and three flux values for (a,b,c): $\phi=0$,  (d,e,f): $\phi=0.2$, (g,h,i): $\phi=0.7$. We choose column: $\Delta=0$, second column: $\Delta=10$ meV, and third column: $\Delta=17$ meV.}
	\label{fig4art}
\end{figure}

Fig. \ref{fig4art} presents the scattering efficiency $Q$ against the  radius $R$ and magnetic field  $B$ for an appropriate choice of the remaining parameters. 
In the absence of $\phi$ and $\Delta$, Fig. \ref{fig4art}a tells us that there is no interaction inside the MGQD below the values $R\approx20$ nm and $B\approx0.4$ T. Beyond that, $Q$ behaves like a wave, forming broad and narrow bands that correspond to specific states of the scattering modes $m$. 
By increasing $\Delta$ in Figs. \ref{fig4art}(b,c), we observe that the interaction within the MGQD starts with smaller values of $B$ compared to the result seen in Fig. \ref{fig4art}a. Furthermore, interaction inside the MGQD occurs at $\Delta=17$ meV even when $B$ is absent. This provides full confirmation of the results presented in Fig. \ref{fig2art}. When the size of the MGQD is below a critical threshold, approximately $R\approx20$ nm, interaction is notably absent. Examining Figs. \ref{fig4art}(d,e,f) for $\phi=0.2$ and Figs. \ref{fig4art}(g,h,i) for $\phi=0.7$, it is evident that clearer bands of $Q$ emerge, signifying the onset of interaction within the MGQD. We conclude that  the interaction within the MGQD becomes discernible at smaller values of $R$ and $B$, with a more pronounced presence in Figs. \ref{fig4art}(g,h,i) at $\phi=0.7$.


\begin{figure}[ht]
	\centering
	\includegraphics[scale=0.45]{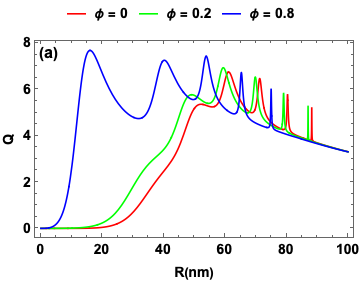}  \includegraphics[scale=0.45]{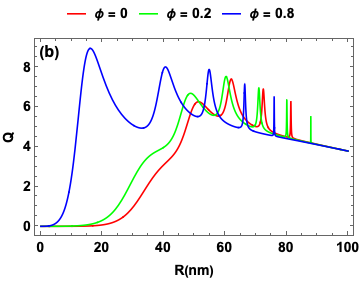}  \includegraphics[scale=0.45]{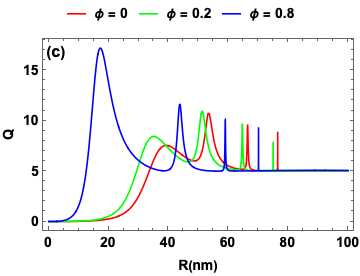}
	\caption{(color online) Scattering efficiency $Q$ as a function of the radius $R=50$ for  $B=1.4$ T,  $E=20$ meV, with (a): $\Delta=0$,  meV for (b): $\Delta=10$,   meV for (c): $\Delta=18$, and  $\phi=0, 0.2, 0.8$, along with (a): $\Delta=0$, (b): $\Delta=10$ meV, (c): $\Delta=17$ meV.}
	\label{fig5art}
\end{figure}

Fig. \ref{fig5art} depicts the scattering efficiency $Q$ plotted against the radius $R$. In the case of $\phi=0$ and $\Delta=0$ in Fig.~\ref{fig5art}a, we observe that $Q$ starts to exhibit non-null values when the radius $R$ approaches $20$ nm, as indicated by the red color. The emergence of the most crucial quasi-bound states is evident only when the radius reaches $R=45$ nm. Unlike the scenario with $\phi=0$, these quasi-bound states initiate their presence at smaller values of $R$ when the flux increases to $\phi=0.2$, as shown by the green color. It is apparent that for $\phi=0.8$ (depicted by the blue color), the scattering efficiency $Q$ starts to exhibit non-null values at smaller values of $R$. Moreover, the resonance peaks associated with these quasi-bound states shift more distinctly to the left, indicating very small values of $R$. Additionally, as the gap widens, the number of excited quasi-bound states decreases. Furthermore, the resonance peaks corresponding to these excited states demonstrate an increase in height, as illustrated in Figs. \ref{fig5art}(b,c).



\begin{figure}[H]
	\centering
	\includegraphics[scale=0.4]{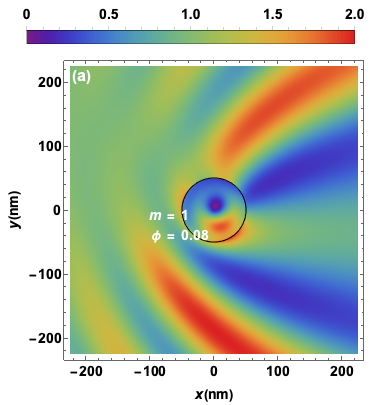}  \includegraphics[scale=0.4]{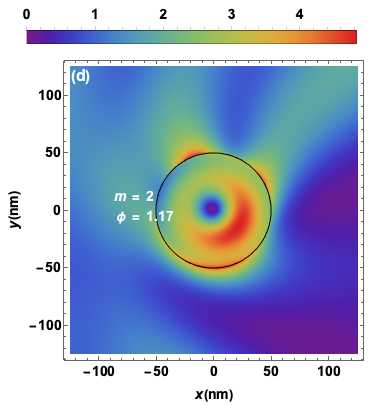}  \includegraphics[scale=0.4]{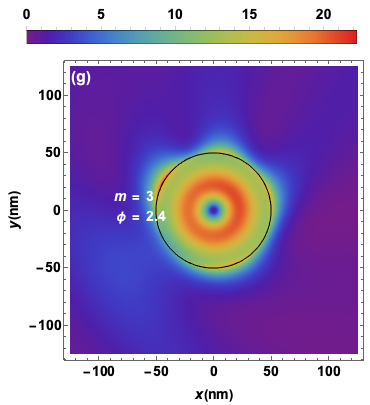}
	
	\includegraphics[scale=0.4]{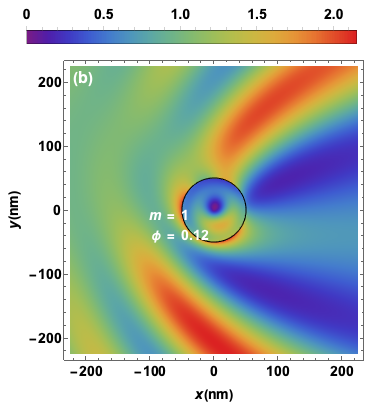} \includegraphics[scale=0.4]{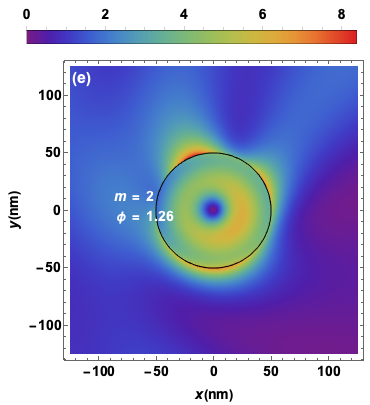}  \includegraphics[scale=0.4]{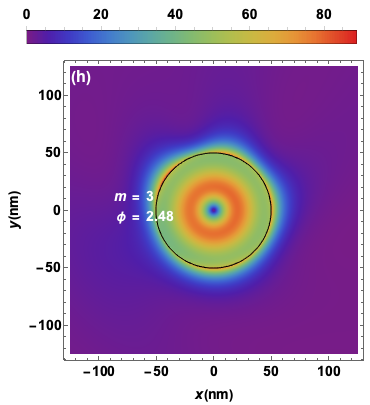} 
	
	\includegraphics[scale=0.4]{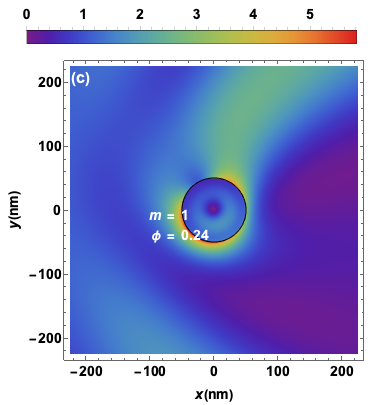} \includegraphics[scale=0.4]{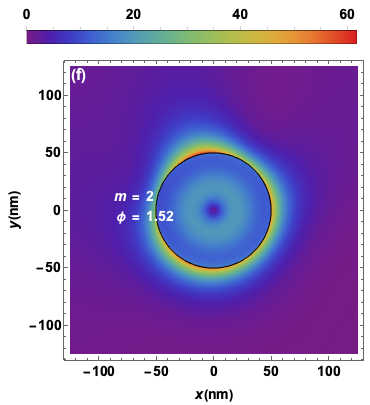} \includegraphics[scale=0.4]{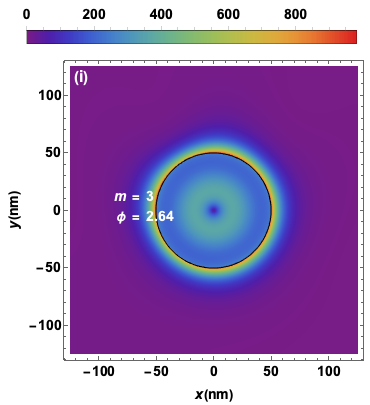} 
	\caption{(color online) The probability density in physical space is illustrated for scattering modes $m=1,2,3$ at $E=20$ meV, $R=50$ nm, and varying values of $\phi$. The gap is set at $\Delta=0,10,17$ meV in the first, second, and third rows, respectively. The MGQD is demarcated by the black circle in the graphical representation.}
	\label{fig6art}
\end{figure}

Fig. \ref{fig6art} depicts the behavior of the probability density $\rho$ in real space for energy $E=20$ meV, and three columns are for modes $m=1,2,3$.
For $m=1$, $\Delta=0$, and $\phi=0.8$, Fig. \ref{fig6art}a indicates a low probability of trapping the electron within the MGQD. 
This is due to the incident electron actively avoiding the MGQD, with the prevailing influence of the diffraction phenomenon serving as the primary explanation for the non-resonant excitation.
On the other hand, we note that a significant part of the density is located on the right-hand side of the MGQD, which is due to the Klein tunneling.
In  Fig. \ref{fig6art}b for $\Delta=10$ meV and  $\phi=0.24$, we observe the diffraction phenomenon is still dominant with a slight decrease in the Klein tunneling.
For $\Delta=17$ meV and $\phi=0.24$, Fig. \ref{fig6art}c shows a complete suppression of the Klein effect and a dampening of the diffraction phenomena.
For $m=2$, $\Delta=0$ and $\phi=1.17$,  Fig. \ref{fig6art}d displays a significant portion of the density concentrated at the boundary and inside the MGQD, which  means that there is a significant chance of keeping the electron inside the MGQD. This explains why there is a non-complete resonance peak, which supports our findings in Fig. \ref{fig3art} (awakening the effects of trapping). 
Furthermore, we observe that a part of $\rho$ at the boundary and center of the MGQD can be improved by increasing $\Delta$ and $\phi$, as clearly seen in Figs. \ref{fig6art}(e,f) for $\Delta=$10 meV, 17 meV, and $\phi= $1.26, 1.52, respectively.
For  $m=3$,  $\Delta=0$, and $\phi=2.4$, Fig. \ref{fig6art}g displays a complete resonance that is shown by the probability density being completely concentrated in the center and near the MGQD. As a result, there is a very high chance of trapping the electron in the MGQD.
By increasing $\Delta$ in Figs. \ref{fig6art}(h,i), it is noticeable that the electron density remains consistently zero outside the MGQD. Simultaneously, there is a distinct enhancement in the density within the MGQD.
From Fig. \ref{fig6art}, we conclude that for excited quasi-bound states with non-resonant peaks, the diffraction phenomenon and Klein tunneling suppress with the increase of gap and AB-flux. Conversely, for excited quasi-bound states displaying full resonance, we note an augmentation in the electron density within the MGQD.


\begin{figure}[H]
	\centering
	\includegraphics[scale=0.46]{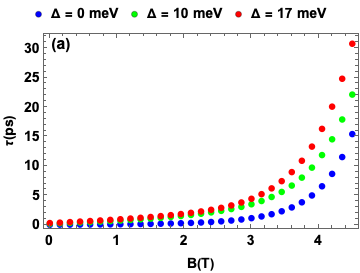}  \includegraphics[scale=0.46]{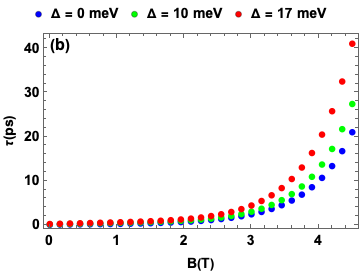}  \includegraphics[scale=0.46]{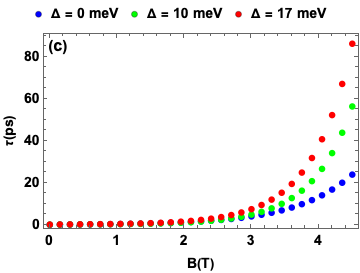}
	\includegraphics[scale=0.46]{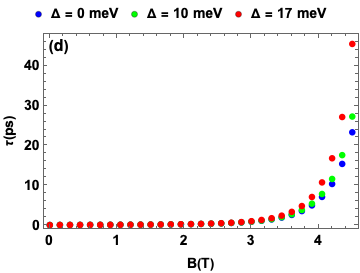}\includegraphics[scale=0.46]{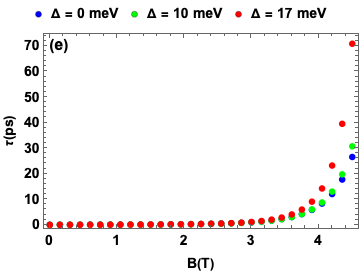}\includegraphics[scale=0.46]{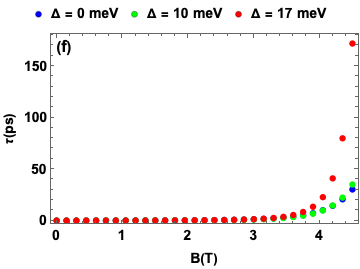}
	\caption{(color online) Lifetime $\tau$ as a function of magnetic field $B$ for $E=20$ meV, and  $R=50$ nm. We choose three values for the gap (blue color):  $\Delta=0$ , (green color): $\Delta=10$ meV, and (red color): $\Delta=17$ meV, along with three values of AB-flux  (a,d): $\phi=0$, (b,e): $\phi = 0.2$, and (c,f): $\phi = 0.8$. The first line corresponds to $m = 1$, and the second one to $m = 3$.}
	\label{fig7art}
\end{figure}

To examine the impact of the flux $\phi$ and gap $\Delta$ on the lifetime of quasi-bound states (trapping time), we treat two specific scattering modes: $m=1$ and $m=3$. Non-resonant excitation leads to weak trapping effects for the first mode, while the second one exhibits full resonance, resulting in strong trapping effects. To accomplish this objective, we will employ numerical analysis of the transcendental equation \eqref{e:34} to ascertain the pairs ($E_i, E_r$) for each excited quasi-bound state, along with the corresponding magnetic field $B$.
Fig. \ref{fig7art} illustrates  the lifetime $\tau$ as a function of the magnetic field $B$, with the first line referring to $m=1$ and the second one to $m=3$. Initially, we observe that trapping electrons in the MGQD is unattainable in the absence of $B=0$. However, with an increase in $B$, the lifetime $\tau$ starts to exhibit non-zero values, indicating the feasibility of electron trapping.
Additionally, it is noteworthy that, in contrast to mode $m=3$, the lifetime for electrons commences at lower values of $B$ for mode $m=1$. This observation aligns with the findings in Fig. \ref{fig2art}, where the quasi-bound state corresponding to mode $m=1$ is excited earlier than that corresponding to mode $m=3$.
In Fig. \ref{fig7art}a for $m=1$, it is evident that without $\phi$ and with an increase in $\Delta$, the lifetime undergoes a substantial improvement. Upon introducing $\phi$, we observe a noteworthy enhancement in the lifetime, reaching approximately $\tau \approx 86$ ps at $\phi=0.8$, as illustrated in Figs. \ref{fig7art}(b,c).
Concerning $m=3$, we  observe that the augmentation of both $\Delta$ and $\phi$ corresponds to a notable enhancement in the duration during which electrons are trapped in the MGQD, as depicted in Figs. \ref{fig7art}(d,e,f).

\section{Conclusion} \label{cc}

We conducted a thorough theoretical investigation of scattering phenomena in a magnetic graphene quantum dot (MGQD) exposed to AB-flux internally and  gap externally. Our approach involved presenting a precise theoretical model that describes the behavior of Dirac fermions and their interaction with a magnetic field, resulting in their confinement within the MGQD structure. To address the problem comprehensively and determine the eigenspinors, we initiated the analysis by solving the Dirac equation analytically. After identifying the appropriate spinors, we applied boundary conditions to derive fundamental formulas for key physical variables essential for studying scattering phenomena suh as the scattering coefficient, scattering efficiency, and probability density. All relevant system parameters, such as the magnetic field, MGQD radius, incident energy, gap, and AB-flux, were considered in formulating these expressions. Subsequently, we extended our analysis into the complex space of incident energies, establishing an equation capable of examining the lifetime of quasi-bound states.


To explore the impact of the gap and the AB-flux on the scattering phenomenon and the lifetime of quasi-bound states, we performed a numerical analysis of the analytical expressions. This involved systematically varying the values of the relevant physical quantities to analyze their effects on the behavior of the system.
We have observed that the scattering efficiency takes significant minimum values even in the absence of a magnetic field. Additionally, we noted that with the increase in gap, the scattering efficiency improves in a remarkable way due to the remarkable broadening of the amplitudes of the resonance peaks corresponding to the excited quasi-bound states.
Conversely, we have found that with an increase in the AB-flux, the probability of generating quasi-bond states at lower magnetic fields becomes more pronounced. Furthermore, we emphasized that with an increase in the AB-flux, the potential for electron scattering into the MGQD becomes feasible at smaller values of the MGQD radius $R$.
By analyzing the probability density for excited quasi-bound states in a non-resonant manner, we observed a damping of the diffraction phenomenon and suppression of Klein tunneling as both the AB-flux and gap  increased. In contrast, for fully resonant excited states, the density at the edge and center of the MGQD experienced a slight increase with the rise in AB-flux and gap. Consequently, the probability of trapping the electron within the MGQD increased.


%


\end{document}